\documentclass[a4paper,fleqn]{cas-dc}

\usepackage[numbers]{natbib}
\usepackage{siunitx}
\usepackage{mathtools}
\usepackage{bm}
\usepackage{tabularx}
\usepackage{booktabs}

\graphicspath{{gfx/}}

\begin{document}
\let\WriteBookmarks\relax
\def\floatpagepagefraction{1}
\def\textpagefraction{.001}
\shorttitle{Multi-detector SHeM}
\shortauthors{C. Zhao et~al.}

\title [mode = title]{A multi-detector neutral helium atom microscope}                      

\author[1]{C Zhao}
\ead{cz390@cam.ac.uk}
\fnmark[1]

\author[1,2]{SM Lambrick}
\ead{sml59@cam.ac.uk}
\fnmark[1]

\author[1]{NA von Jeinsen}

\author[1]{Y Yuan}

\author[1]{X Zhang}

\author[1]{A Radi\'{c}}

\author[1,2]{DJ Ward}

\author[1]{J Ellis}

\author[1]{AP Jardine}

\affiliation[1]{organization={Department of Physics, University of Cambridge},
	addressline={19 J.J. Thomson Avenue}, 
	city={Cambridge},
	postcode={CB3 0HE},
	country={United Kingdom}}
	
\affiliation[2]{organization={Ionoptika Ltd.},
                addressline={Units B5-B6, Millbrook Close}, 
                postcode={S053 4BZ},
                city={Chandlers Ford},
                country={United Kingdom}}

\cortext[cor1]{Corresponding author}
\fntext[fn1]{These two authors contributed equally.}

\begin{abstract}
Scanning helium microscopy (SHeM) is an emerging technique that uses a beam of neutral atoms to image and analyse surfaces.  The low energies ($\sim$64 meV) and completely non-destructive nature of the probe particles provide exceptional sensitivity for studying delicate samples and thin devices, including 2D materials.  To date, around five such instruments have been constructed and are described in the literature.  All represent the first attempts at SHeM construction in different laboratories, and
use a single detection device. Here, we describe our second generation microscope, which is the first to offer multi-detector capabilities. The new instrument builds on recent research into SHeM optimisation and incorporates many improved design features over our previous instrument. We present measurements that highlight some of the unique capabilities the instrument provides, including 3D surface profiling, alternative imaging modes, and simultaneous acquisition of images from a mixed species beam.

\end{abstract}

\begin{highlights}
\item The first multi-detector helium microscope is presented
\item Significant usability and modularity benefits are built into the design
\item 3D profilometry is possible with no increase in measurement time
\item Imaging with multiple neutral species is demonstrated for the first time
\item A broadened range of contrast mechanisms thanks to simultaneous imaging modes
\end{highlights}

\maketitle
\section{Introduction}
\label{sec:int}

Scanning helium microscopy (SHeM)\cite{witham_simple_2011,palau_neutral_2023,barr_design_2014,bhardwaj_neutral-atom-scattering-based_2022} is an emerging  technique for surface characterization that uses a narrow beam of neutral helium atoms. The helium beam can be scanned over a surface to produce maps of scattered helium flux which can be rendered as micrographs, or can be used to perform spot-profile analysis, e.g. through diffraction \cite{von_jeinsen_helium_2024}. The use of neutral helium atoms as the probe particles has some unique benefits, which makes SHeM especially suitable for the study of insulators, optically transparent materials, ultrathin or 2D materials\cite{von_jeinsen_helium_2024,radic_defect_2024} and other delicate samples. Key to all these advantages is the exceptionally low energy of the incident helium atoms, typically $\sim\SI{64}{\milli eV}$ for a room temperature beam, in contrast to electron microscopy where at least $\sim\SI{100}{eV}$ is normally required\cite{joy_choosing_2009}. Using significantly more massive particles than electron microscopy also means that the de Broglie wavelength of the helium atoms, for a room temperature beam, is $\SI{0.05}{\nano\metre}$\cite{bracco_surface_2013}, enabling scattering experiments to provide atomic scale information and putting no fundamental physical restriction on the resolution with current technology. The extremely low energy makes SHeM exclusively surface sensitive; the classical turning point for thermal helium atoms at a surface is $2-\SI{3}{\angstrom}$ from the cores of the top-layer atoms\cite{farias_atomic_1998}.  Similarly, the neutral and chemically unreactive probe species means there is no need for special sample preparation, such as ensuring conductivity or surface evenness, as is often needed in electron microscopy or scanning probe microscopy.

Several different methods for producing a narrow beam of atoms have been proposed in the literature, which include focusing the beam with atom mirrors\cite{holst_atom-focusing_1997} and focusing with Fresnel zone plates\cite{eder_focusing_2012,flatabo_reflection_2024}; both of these are extremely challenging.  Hence, in the current work we employ collimation by a micron-scale pinhole\cite{barr_design_2014,witham_simple_2011}, as the most straightforward approach and the one most commonly used in the literature.

\begin{figure}
	\centering
	\includegraphics[width=0.5\textwidth]{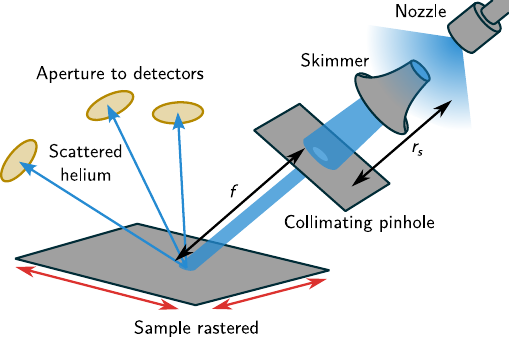}
	\caption{Schematic depicting the SHeM technique. A helium beam is produce by selecting the centreline from supersonic expansion using a skimmer; the beam is further collimated to a microscopic size by a pinhole. After scattering on the sample, the helium atoms through the detector apertures are collected by the detectors. Rastering of the sample under the beam results in a helium micrograph. The working distance, $f$, and the skimmer-pinhole distance, $r_s$, are annotated.}
	\label{fig:SHeM Tech}
\end{figure}

Figure \ref{fig:SHeM Tech} demonstrates the principle of SHeM via a simplified schematic of the instrument presented in the current work. First, a high-intensity helium beam is formed by a supersonic expansion\cite{scoles_atomic_1988,barr_desktop_2012,eder_velocity_2018} through a nozzle. The nozzle typically has a $5-\SI{10}{\micro\metre}$ diameter and is connected to a $\sim \SI{100}{bar}$ reservoir of helium gas. The helium beam is then collimated by a cone-shaped beam dynamics skimmer, with a $\SI{100}{\micro\metre}$ diameter aperture, and further by a $0.3-\SI{30}{\micro\metre}$ diameter pinhole placed near the sample. The `source-distance', $r_s$, between the source and pinhole, and the `working-distance', $f$, between the pinhole and sample, are both crucial and are discussed in detail below. Subsequently, the collimated helium beam scatters from the sample surface and the detector(s) collect scattered atoms that pass through several detector-defining apertures.  The helium flux is detected using either a quadrupole mass-spectrometer or a custom high sensitivity helium detector\cite{bergin_low-energy_2021}. To create the image, the sample is rastered beneath the collimated beam using a nanoscale-precise scanning stage, to generate a scattered helium map which may be presented as an image.

To date, SHeM has been applied widely in measurements ranging from characterising atomically thin films\cite{bhardwaj_neutral-atom-scattering-based_2022,bhardwaj_contrast_2023,eder_sub-resolution_2023,von_jeinsen_helium_2024,radic_defect_2024} to examining microscopic biological structures\cite{myles_taxonomy_2019,lambrick_multiple_2020,fahy_highly_2015}. However, all have relied on a single detector\cite{barr_design_2014, koch_imaging_2008, witham_simple_2011, flatabo_reflection_2024, bhardwaj_neutral-atom-scattering-based_2022}, acquiring a single data point at a time, at only modest rates. One particular application, which is of growing interest but where linear data acquisition is particularly limiting, is 3D reconstruction of surface topography.  Two alternative reconstruction methods have been proposed in the literature\cite{radic_3d_2024,lambrick_true--size_2021,myles_taxonomy_2019}, but both rely on multiple helium images.  Conventionally, these must be acquired sequentially, significantly increasing total acquisition time and making reconstruction vulnerable to drift and difficulties with subsequent image registration. Similarly, the emergence of new contrast mechanisms and approaches such as atom micro-diffraction\cite{von_jeinsen_2d_2023,hatchwell_measuring_2024} demonstrate the importance of making simultaneous measurements under different scattering conditions.

In the current work we present the first multi-detector helium atom microscope, the Cambridge `B-SHeM', which has the capacity for 4 detectors, as well as significant usability improvements over the first generation instrument. We start by outlining design considerations for the instrument and giving details of the implementation of those considerations. After describing the instrument itself, we explore the several applications of multi-detector helium microscopy and use selected measurements as case studies.

\section{Instrument Design}



\subsection{Guiding Principles}

The underlying principles of the Cambridge B-SHeM design were to develop a compact, self-contained instrument with a relatively small footprint, while also optimising performance and usability in light of experience gained from use of first generation SHeMs.  In particular, we have experience with the Cambridge “A-SHeM” and the Newcastle SHeM, both of which follow the design presented by Barr et al. \cite{barr_design_2014}.  Our new instrument is more comparable with a typical SEM in footprint than the previous SHeM instruments, or indeed any other helium atom scattering (HAS) instruments.  Our approach also simplifies much of the practical implementation. Several key design decisions were made at an early stage, which included:

\begin{itemize}
	
	\item \textbf{Establishing a multidetector geometry} to enable simultaneous parallel measurements at different scattering conditions, giving new functionality and improving both performance and resilience.  The design also allows both commercial and bespoke detectors to be utilised at the same time.
	
	\item \textbf{Mounting the sample in a horizontal orientation}, to avoid the difficulties associated with sample adhesion and manipulation of a vertically mounted sample, resulting in an instrument with a vertical scattering plane unlike previous SHeM and most HAS implementations.
	
	\item \textbf{Choosing a nominal incidence angle of 30$^\circ$} for the helium microprobe on the sample, as a compromise between achieving best resolution at normal incidence \cite{bergin_method_2019} and the benefit of larger non-normal incidence angles enabling measurements of specular and enhancing topographic contrast.  Previous generations of SHeM have used a 45$^\circ$ incidence angle \cite{barr_design_2014, bhardwaj_neutral-atom-scattering-based_2022} or normal incidence \cite{witham_simple_2011, flatabo_reflection_2024}.
	
	\item \textbf{Optimising the atom-optical geometry} of the system in a flexible way, to maximise signal, enabling sub-micron resolution to be achieved, and to allow the pinhole collimation to be upgraded to Fresnel zone-plate focussing for higher resolution.
	
	\item \textbf{Aiming to use a working distance, $f$, of up to about 1\,mm.} $f$ is the normal distance between the sample and pinhole, and choosing such values enables a wide range of samples to be studied, include those with topographic structure on the mm scale.
	
	\item \textbf{Capability for sample transfer and sample preparation} to be added as required in the future.  These are not straightforward to incorporate in existing SHeM sample chambers.
	
\end{itemize}

In the following sections we describe our overall instrument design, which achieves these goals, followed by giving further relevant details on each of the main components.

\begin{figure*}
	\centering
	\includegraphics[width=0.8\textwidth]{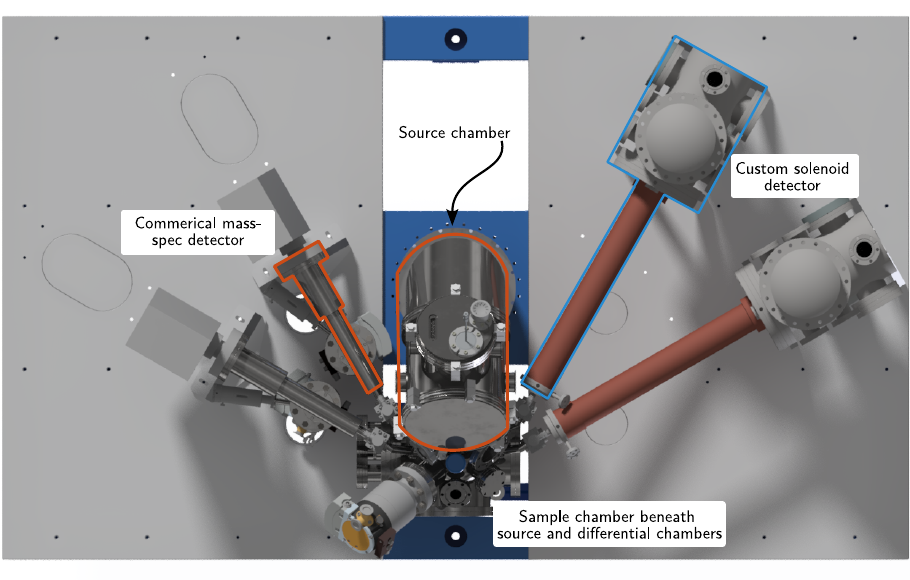}
	\caption{A top-down render of the Cambridge B-SHeM, where the detectors arranged around the central sample chamber, can be seen. On the left are two commercial RGA mass spectrometers, and on the right are two custom solenoid detectors. The Sample chamber is located beneath the source chamber, as can be seen in figure \ref{fig:bshem_cross_section}.}
	\label{fig:top_down}
\end{figure*}

\subsection{Overall Instrument Design}

Figure \ref{fig:top_down} shows a top-down schematic of the B-SHeM instrument.  The main table (2\,m $\times$ 1.2\,m) holds the helium source and sample chamber (middle), surrounded by up to four detectors.  The table size was adjusted to provide sufficient space for either commercial quadrupole mass-spectrometer detectors (left), or instances of the fifth generation version of an in-house high sensitivity solenoidal/magnetic sector detector (right). The table itself is constructed of 100\,mm box section steel, designed with sufficient strength to withstand the hard shutdown torque of the large source turbomolecular pump that is required, which is mounted below the table for stability.  The table is then attached, using passive vibration isolation bellows (and constrained using high-strength motion limiting bars), to a similarly constructed support frame which also contains 19” racking to house the control electronics.  Sufficient space on the main table is allowed to the left and right of the sample chamber, such that the sample manipulator can be removed or further developed, and so that the sample transfer system can be added, respectively.

\begin{figure*}
	\centering
	\includegraphics[width=0.75\linewidth]{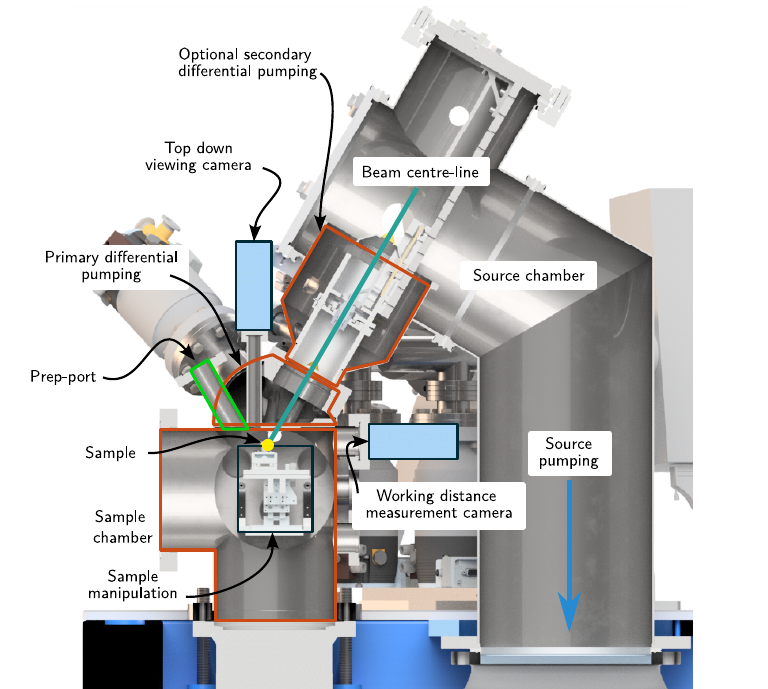}
	\caption{A cross-sectional render of the B-SHeM, with the sample location, sample manipulation, and incident beam line highlighted. The detectors are connected out of plane of the render, and can be seen in figure \ref{fig:top_down}. Two optical view ports are highlighted in blue, one to measure the working distance, and the other to allow optical targeting of the sample. The preparatory port, highlighted in green, provides line of sight from a standard CF flange to the sample area allowing for \emph{in situ} sample preparation.}
	\label{fig:bshem_cross_section}
\end{figure*}

Figure \ref{fig:bshem_cross_section} shows a cross-section through the scattering plane of the instrument. The helium beam is initially formed in a free-jet expansion at the nozzle inside the source chamber (middle top, discussed), which points downwards at an angle of 30$^\circ$ to the vertical.  The beam then passes through a skimmer and through one differential pumping stage into the sample chamber (lower left).  The sample and differential chamber form one integrated assembly (left), which is mounted directly on the table.  The beam is collimated by a micron-sized pinhole mounted in the “pinhole-plate”, which is fitted to the separator between the differential and sample chambers, and selects the scattered helium that is transferred to the detectors.

\subsection{Source Geometry}

Careful design of the source atom-optical geometry is crucial to maximising intensity, and thus the ability to achieve high resolution.  Several articles have been published discussing different approaches to optimising SHeM design\cite{salvador_palau_theoretical_2017,bergin_method_2019,palau_theoretical_2016}. The B-SHeM is designed to enable the use of a large range of source geometries, including those comparable to the first generation A-SHeM, as well as geometries optimised for high resolution imaging and spot profile measurements. The geometry most dissimilar to those used in the A-SHeM are likely to be future high resolution imaging, therefore a summary of the an optimisation used to inform the design is given below.

As discussed by Bergin et. al \cite{bergin_method_2019}, the lateral size of the helium beam at the sample has contributions due to the pinhole size, the source angular spread, and diffraction at the pinhole.  Approximating each term as a Gaussian distribution gives the following expression \cite{bergin_method_2019} for the size of the resulting spot on the sample, as a standard deviation, $\phi_p$,
\begin{equation}\label{eq:optimisation_eq}
	\phi_p=\sqrt{ \left( \frac{d}{2\sqrt{3}} \right)^2 + \left( \frac{\beta f}{\sqrt{3}} \right)^2 + \left( \frac{0.42 \lambda f}{d} \right)^2  }.
\end{equation}
Here, $d$ is the aperture size, $\beta$ is the source angular spread, $\lambda$ is the helium wavelength, and $f$ is the working distance. Performance of the microscope is maximised when the terms are equal in magnitude, so by equating the first and third, an optimum condition can be determined for a given wavelength and working-distance. For a room temperature beam and a $\SI{1}{\milli\metre}$ working-distance, a beam size of $\sim\SI{450}{\nano\metre}$ is optimal, while for a $\SI{0.25}{mm}$ working distance, that optimal beam size reduces to $\sim\SI{230}{\nano\metre}$. To maximise intensity, the source angular spread, $\beta$, must then be adjusted to match this optimum.

In the free-jet nozzle expansion, gas appears to diverge from a `virtual-source”'of a finite size, positioned in front of the nozzle.  A sufficiently small source angular spread can be achieved with either a large virtual-source and corresponding large `source-distance', $r_s$ (see the geometry in Fig. \ref{fig:SHeM Tech}), or a smaller virtual-source and shorter source-distance. Depending on the nozzle operating conditions and the local scattering processes, the virtual-source size can vary between a size close to that of the nozzle (usually $\sim\SI{10}{\micro\metre}$), and a much larger size that is controlled by the size of the skimmer (typically $\sim\SI{100}{\micro\metre}$).  In practise, conditions nearer the latter are usually found to maximise intensity during measurements\cite{bergin_instrumentation_2018}.

Given the requirement for a $\sim\SI{1}{\milli\metre}$ working distance and a skimmer limited virtual source size of $\sim\SI{100}{\micro\metre}$, achieving a beam-size of $\sim\SI{500}{\nano\metre}$ would require a source distance of $\sim\SI{300}{\milli\metre}$ (i.e. the source size must be `demagified' by a factor of $\sim300$). If a lower resolution can be accepted the source-distance can be reduced giving an increase in the signal level, and similarly if smaller high brightness sources can be achieved \cite{bergin_method_2019} the source-distance can be reduced. For resolutions comparable to the published data from the A-SHeM\cite{lambrick_multiple_2020} source distances in the range $50-\SI{100}{\milli\metre}$ could be used. Because of range of different operating conditions and the uncertainties in future source improvement, the B-SHeM was designed to be configurable in a range of different states, with source-distances from $\sim\SI{55}{\milli\metre}$ to $\sim\SI{300}{\milli\metre}$. 

\begin{figure*}
	\centering
	\includegraphics[width=0.6\linewidth]{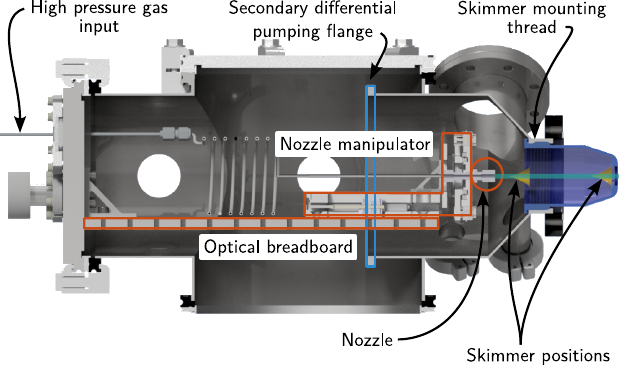}
	\caption{A schematic of the internals of the source chamber, with two possible skimmer positions rendered. The skimmer mount that threads into the chamber is presented in translucent blue. The optional secondary differential pumping chamber can be isolated at the location of the highlighted ISO flange.}
	\label{fig:bshem_source}
\end{figure*}

In Fig. \ref{fig:bshem_cross_section}, the B-SHeM source chamber is shown pointing diagonally downwards, while Fig. \ref{fig:bshem_source} shows more internal detail.  The large source vacuum pump is a crucial component, as it maintains the low pressure around the free-jet expansion. Although helium beam sources typically use pumps in excess of 2000\,L/s, test measurements on the Cambridge A-SHeM suggested that little degradation in SHeM performance would result from reducing the source pumping speed to about 1200\,L/s.  Our long term intention is to use smaller nozzles at higher pressure, to achieve greater brightness with less gas throughput, also meaning a smaller source pump can be used. A \emph{Leybold MAG W 1300} was therefore selected, particularly given its performance under high gas throughput.  It is connected to the source chamber through a large diameter pipe that provide a high gas conductance to maintain pumping speed near the nozzle.  

The skimmer is standardly fixed to a mounting assembly that screws into the large thread at the end of the chamber (right); different length mountings enable the skimmer to be positioned at various places inside the source chamber or, if small source-distances are required, protruding into the differential chamber (as shown).  Alternatively, for long source-distances, the skimmer can be mounted from the internal flange (middle).  Windows are positioned to enable the tip of the skimmer to be observed at source-distances of $\sim\SI{300}{\milli\metre}$, $\sim\SI{200}{\milli\metre}$ and $\sim\SI{120}{\milli\metre}$, for nozzle alignment, making these particular source-distances preferred options.  When the skimmer is mounted from the internal flange, the differential pumping can optionally be split into two separately pumped stages, using the additional DN63CF port (upper right).

The nozzle assembly, shown in the centre of Fig. \ref{fig:bshem_source}, is mounted on a custom manipulator consisting of 3 piezoelectric drives and encoders. The use of an encoded manipulator makes nozzle alignment and adjustment of the nozzle-skimmer distance straightforward, and assists with the characterisation of source performance. For large changes in the axial nozzle position, the manipulator position on an optical breadboard can be adjusted.

\subsection{Differential Pumping \& Sample Environment }

In order to maintain a low background during measurements, differential pumping is required between the source and sample environment.  The sample and differential pumping chambers are fully integrated, as this means the requirement for small source-distances can be more easily achieved compared to using separable chambers.

Figure \ref{fig:bshem_cross_section} shows the combined chamber in cross-section.  The sample chamber consists of a 150\,mm diameter 5-way cross, which can be accessed from the front or side to load samples.  The body of the chamber provides good conduction of background gas from the sample region to a dedicated $\sim$400\,L/s turbopump mounted below.  Differential pumping was achieved by attaching a separate hemispherical chamber to the top of the cross, which connects to the source chamber.  The plate that forms the boundary between the sample and differential chambers holds the pinhole plate.  Sealed connecting tubes between the pinhole plate and detectors pass through the differential pumping region towards the detectors, that are positioned out of plane of the render.  Further ports pass through the differential chamber to the sample chamber, one at normal incidence, which is intended to be used with an optical targeting camera for coarse sample positioning, and a non-normal port intended for sample preparation.  A second optical viewport is positioned on the side of the sample chamber, which is intended to be used with an additional camera to provide non-contact measurements of the working distance; both camera ports are indicated in blue in Fig. \ref{fig:bshem_cross_section}.

Sample manipulation is achieved with a 4-axis set of piezoelectric manipulators from the \emph{SLS-23 UHVT} series from \emph{SmarAct}. From top to bottom the manipulation stack consists of an azimuthal rotation stage, a short range linear motor, $x$, a long range linear motor, $y$, and a high force lift stage, $z$. The use of a high force piezo for the $z$ stages allows for more complex sample environments, such as heating and cooling, and/or much larger samples than previously accessible.  The $y$ stage can move the sample between the helium imaging position, the position below the viewport for optical targeting and the position below the sample preparation port.  The long travel in the $y$ and $z$ directions also allow the sample to be moved well away from the pinhole during sample exchange.

Multiple ports provide access to the beam-axis in both the source and differential pumping chambers, allowing insertion of intermediate apertures if required.  Such apertures could be used to control the effective source size in situ, enabling rapid changes in resolution\cite{flatabo_fast_2018} and to reduce background due to stagnation of gas behind the pinhole\cite{fahy_highly_2015,sam_m_lambrick_formation_2021}. The installation of a slit aperture ($0.5\times \SI{6}{\milli\metre}$) on the beam axis in the primary differential pumping chamber, seen near the centre of figure \ref{fig:bshem_cross_section}, should prevent excess helium stagnating behind the pinhole. We have found the aperture to reduce background observed in helium micrographs by $\sim38\%$.

\subsection{Modular pinhole-plate}

The modular `pinhole-plate', which defines the scattering geometry of the instrument, is a particularly successful feature of the A-SHeM microscope \cite{barr_design_2014}, which we have further developed here to include multiple detectors.  The new pinhole-plate attaches to the top of the sample chamber/bottom of the differential chamber, and makes sealed connections with both the beam-source and detectors.  It incorporates both the incident-beam defining pinhole and the apertures which define the detection conditions in a single assembly, so it can be constructed with high precision, as needed for the required working distances.  The pinhole-plate can easily be swapped for different measurements in different SHeM operating configurations.

\begin{figure}
	\centering
	\includegraphics[width=\linewidth]{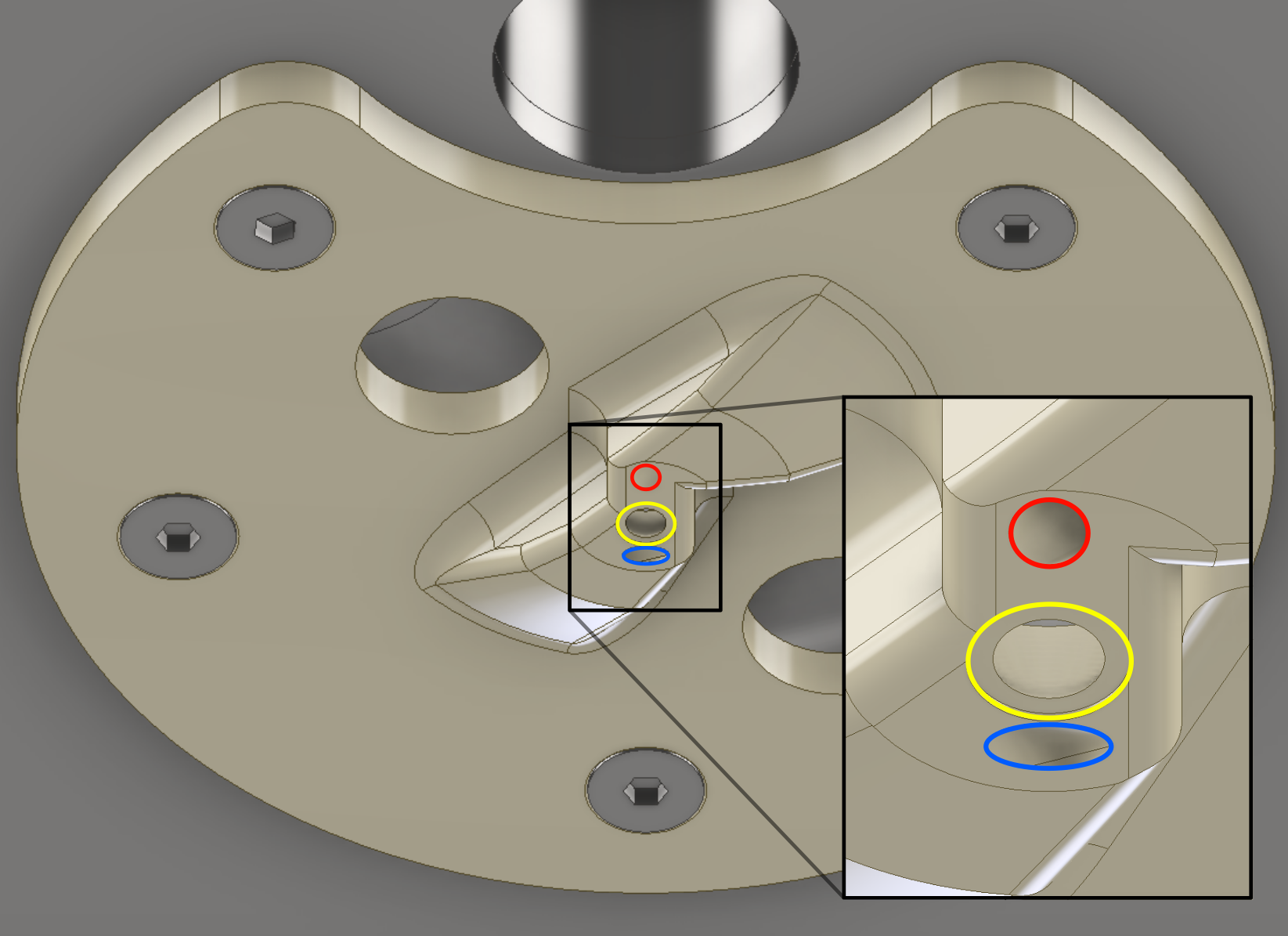}
	\caption{A render of the pinhole plate installed in the sample chamber. This pinhole plate has two detector apertures, one for forwards scattering highlighted in red, and one for back scattering highlighted in blue; two holes for unused detectors can also be seen. The pinhole would be mounted in the yellow highlighted opening. The pinhole-plate is mounted directly to the chamber with five screws, one is behind the inset, allowing easy swapping of the plate whenever the sample chamber is vented.}
	\label{fig:pinhole_plate}
\end{figure}

An example pinhole-plate, configured to use two detectors, is shown schematically in Fig. \ref{fig:pinhole_plate}.  Here, the pinhole aperture, usually produced by focussed ion-beam milling a silicon nitride membrane, is mounted in the recess indicated in yellow.  The detector apertures marked in red and blue collect forward and backward scattered atoms, and direct them to the detector connections through the internal structure of the plate (upper right, lower left).  The two open apertures through the plate (upper left, lower right) connect the unused detector ports directly to the sample chamber.  The region around the apertures is raised relative to the rest of the pinhole-plate, to maximise background gas conduction away from the sample region.

More generally, the positioning of detection apertures is only limited by manufacturing methods, and by varying the internal structure of the pinhole plate, atoms collected in any collection aperture can be transferred to any detector port, depending on the requirements of a particular experiment. As demonstrated by work on our previous instrument, different pinhole plates allow completely different imaging modes to be accessed, for example by changing the incidence angle of the machine \cite{radic_3d_2024}, or by enabling high angular resolution for reciprocal space measurements \cite{von_jeinsen_2d_2023}.  The use of 3D printed components, including both metals and vacuum compatible SLA resin \cite{radic_application_2023}, has enabled rapid prototyping of different pinhole-plate designs, and the exploration of geometries difficult to achieve with conventional manufacturing.

\subsection{Detection}

As SHeM detection operates through stagnation \cite{bergin_instrumentation_2018} direct line-of-sight is not needed between the sample and the detector ionization volume, as is often required in traditional macro-scale  helium atom scattering instruments. Consequently, the detectors can be placed around the sample chamber, with the effective detector position with respect to the scattering centre controlled by the pinhole-plate.

Four detector connections are included in the present design, which pass through internal tubes from behind the pinhole plate in the sample chamber, through the principle differential pumping chamber, to external flanges just outside the chamber assembly. Subsequently, DN16CF gate valves are used to provide isolation between the detectors and sample chamber, and the bore of the connecting pipework is adjusted to be sufficiently large to avoid limiting conductance between the collection apertures and detector. The system can be used with both commercial mass spectrometers or custom high sensitivity helium detectors \cite{bergin_low-energy_2021}, or a combination of both. 

Data in the current work has been acquired using \emph{Hiden HAL/3F RC301 PIC300} residual gas analysers arranged in a stagnation configuration inside a detector chamber with a volume of approx. 1\,L.  Each detector is equipped with a throttle valve, consisting of a moving vane that controls a diamond shaped aperture, enabling the pumping speed to be varied between about 0.1\,$L\,s^{-1}$ and 50\,$L\,s^{-1}$.  The open area varies quadratically with the vane position, thus giving accurate control of pumping speed at small openings. Variable detector pumping can be used to tune the trade-off between efficiency and detector response time, as discussed by Myles et al. \cite{myles_fast_2020}.

\subsection{Control System}

The whole microscope is controlled using a distributed computer control system, connected over a local laboratory ethernet.  Microcontrollers and Raspberry Pi single board computers are used to interface with hardware directly, generally receiving commands and transmitting data using the \textsc{mqtt} protocol -- a lightweight publish/subscribe messaging protocol for the Internet of Things (IOT)\cite{noauthor_mqtt_nodate,quincozes_mqtt_2019,pahomqtt_2024}.  High level measurement scripts are run on a Windows PC, with scripting of SHeM experiments and live data visualisation being performed in \textsc{matlab}.  The approach removes a single point of failure for the instrument, since multiple redundancies can be put in place, mediated only by the network.  It also makes it particularly to integrate new hardware into the system, often just requiring a single connection to the instrument PoE switch. We use Raspberry Pi HQ camera modules on the optical viewports, with associated Raspberry PIs, through the supplied web-brower interface. To enable fast measurements, with effectively zero dead-time between counting bins, a custom, fast, counting system has been developed, providing timestamped helium counting with gating time precision of $\pm 4\,\mathrm{\mu s}$.

\section{Applications}
\label{sec:applications}

We now give case studies of several new images modalities made possible by the new instrument.  We cover simultaneous scattering measurements, parallelisation of 3D reconstruction, and concurrent measurement of mixed species in a single beam.

\subsection{Simultaneous imaging modes}
\label{sec:multiple_modes}

The majority of SHeM data obtained to date has used `forwards-scattering' detection, where the momentum transfer between the incident and scattered helium is mostly in the direction normal to the surface. Focusing on forwards scattering is a pragmatic choice, where one must be made, to allow both topographical measurements and studies of the specular or first order diffraction channels. However, the availability of multiple detectors allows us to simultaneously measure at multiple scattering conditions, as well as reducing  measurement time.  Simultaneous, parallel measurements also ensure the sample is in precisely the same condition for each.

\begin{figure}
	\centering
	\includegraphics[width=\linewidth]{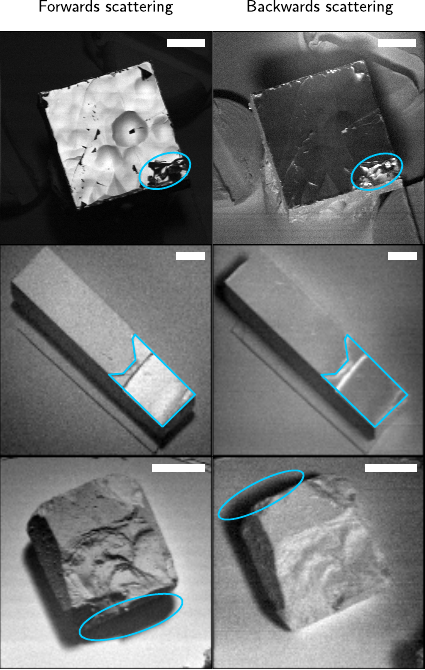}
	\caption{Three samples imaged with both forwards and backwards scattering modes simultaneously. From top to bottom, the samples were: heated HOPG, cleaved LiF, and a sugar crystal. The HOPG was measured using the pinhole plate shown in figure \ref{fig:pinhole_plate} while the LiF and sugar were measured using the design presented in figure \ref{fig:3D_method}. All scale bars \SI{1}{\milli\metre}.}
	\label{fig:forwards_backwards_scattering}
\end{figure}

To demonstrate the capabilities of simultaneous imaging modes, the instrument was configured for forwards and backwards scattering detection.  The forwards scattering detection was arranged to be similar to previous SHeM configurations. Backwards scattering detection has not previously been measured and involves detected the helium that is scattered back in the approximate direction of the source. Figure \ref{fig:forwards_backwards_scattering} shows three samples imaged with both forwards and either full backwards scattering or partial backward scattering. The first is a sample of HOPG that has been heated to remove adsorbates; it exhibits partially ordered scattering, having most of the intensity scattered approximately in the direction of the specular. We see that the sample is significantly brighter in the forwards scattering, while it is dark in the backwards scattering micrograph.  Small deviations in the surface are emphasised in the forwards scattering, while the back scattering mostly captures large scale topography.  The exceptions are a few bright flakes in the bottom right of the sample, these same flakes appear dark in the forwards scattering micrograph. The second sample was a cleaved LiF crystal, exhibiting ordered scattering. We see two domains on the LiF; one (highlighted in figure \ref{fig:forwards_backwards_scattering}) shows detection of a diffraction channel in the forwards scattering detector and thus appears bright, while appearing dark in the backwards scattering detector, similar to the HOPG. The other domain, where no diffraction channels are directed into either detector, renders similarly dull in both images. The third sample was a sugar crystal, demonstrating disordered diffuse topographic contrast dominated by masking; it can be seen that by swapping from forwards to backwards scattering the mask, highlighted in blue in figure \ref{fig:forwards_backwards_scattering}, swaps sides of the crystal, always seen opposite to the direction of detection. 

\subsection{3D imaging}

Heliometric stereo\cite{lambrick_true--size_2021,radic_3d_2024} is a method for performing 3D surface profilomtry using helium atoms, that relies on a known model for the scattering of helium from disordered/`technological' surfaces\cite{lambrick_observation_2022}. The approach is an adaptation of photometric stereo\cite{woodham_photometric_1980}, an established approach used widely with visible light, to helium microscopy. The method uses helium micrographs collected with different detectors, equivalent to different illumination angles with light, to produces a topographic map. Our previous work\cite{radic_3d_2024} has demonstrated heliometric stereo with a single detector instrument using sample manipulation in place of multiple detectors, however, that approach comes with a number of drawbacks, including extended acquisition times as many helium micrographs must be acquired sequentially, and decreased effective resolution due to the necessity of semi-manual image alignment. In the current work, we use three detectors to directly perform heliometric stereo. We outline a brief summary of the formalism of heliometric stereo below, further details and derivations can be found in Lambrick \& Salvador-Palau et al.\cite{lambrick_true--size_2021}. 

\begin{figure}
	\centering
	\includegraphics[width=\linewidth]{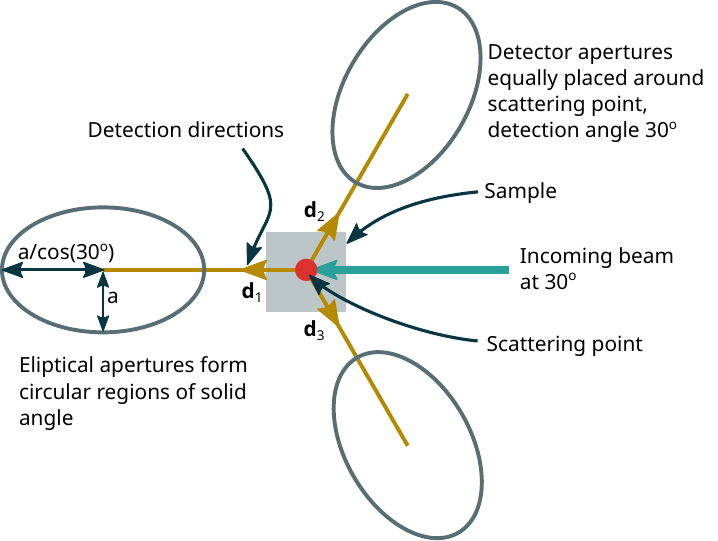}
	\caption{Top-down geometric setup of the 3D profilometry pinhole-plate, 3 detectors are placed equally spaced around the scattering point while the incident beam is at \ang{30} to the surface normal.}
	\label{fig:3D_method}
\end{figure}

The dominant scattering mechanism from a disordered/ `technological' surface is diffuse scattering\cite{lambrick_observation_2022}, which may be modelled with a cosine distribution, allowing us to write the detected intensity as
\begin{equation}\label{eq:cosine}
	I(\theta) = \rho\cos\theta = \rho\hat{\bm{n}}\cdot\hat{\bm{d}},
\end{equation}
where $\hat{\bm{n}}$ is the local surface normal and $\hat{\bm{d}}$ is a vector pointing from the sample to the detector. For the current work, the detection directions are shown in figure \ref{fig:3D_method}. Equation \ref{eq:cosine} may be expressed as a matrix equation
\begin{equation}\label{eq:ch3D:basic_photostereo}
	\Vec{I}_{(x',y')}=\rho \mathsf{D}\hat{\bm{n}},
\end{equation}
where $\Vec{I}$ is a vector of pixel intensities for different detection direction. The matrix $\mathsf{D}$ contains the  vectors $\hat{\bm{d}}_i$, these are shown in figure \ref{fig:3D_method}. Equation \ref{eq:ch3D:basic_photostereo} may be solved thus,
\begin{gather}
	\rho_{(x^\prime,y^\prime)} = |\mathsf{D}^{-1}\Vec{I}_{(x^\prime,y^\prime)}|,\label{eq:ch3D:solvePhotostereo1}\\
	\hat{\bm{n}}_{(x^\prime,y^\prime)} = \frac{1}{\rho_{(x^\prime,y^\prime)}}\mathsf{D}^{-1}\Vec{I}_{(x^\prime,y^\prime)}.\label{eq:ch3D:solvePhotostereo2}
\end{gather}
Finally, provided the height of the surface can be described by a continuous function, {\em i.e.} $z = f(x, y)$, then
\begin{gather}
	\hat{\bm{n}}(x,y) = \bm{\nabla} F(x,y,z) = \bm{\nabla}[z- f(x,y)],\label{eq:ch3D:gradient_field}
\end{gather}
which can be integrated to give the surface profile, $z = f(x,y)$. A regularized least squares approach, developed by Harker and O'leary\cite{harker_least_2008,matthew_harker_surface_2021}, is used in the current work.

To implement heliometric stereo on the Cambridge B-SHeM we manufactured a pinhole-plate with 3 equally spaced detector apertures, each occupying the same sized circular areas of solid angle.  A schematic of the configuration of detection is shown in figure \ref{fig:3D_method}, where there is a single forwards scattering detector and two partially backwards scattering detectors. 

\begin{figure}
	\centering
	\includegraphics[width=0.85\linewidth]{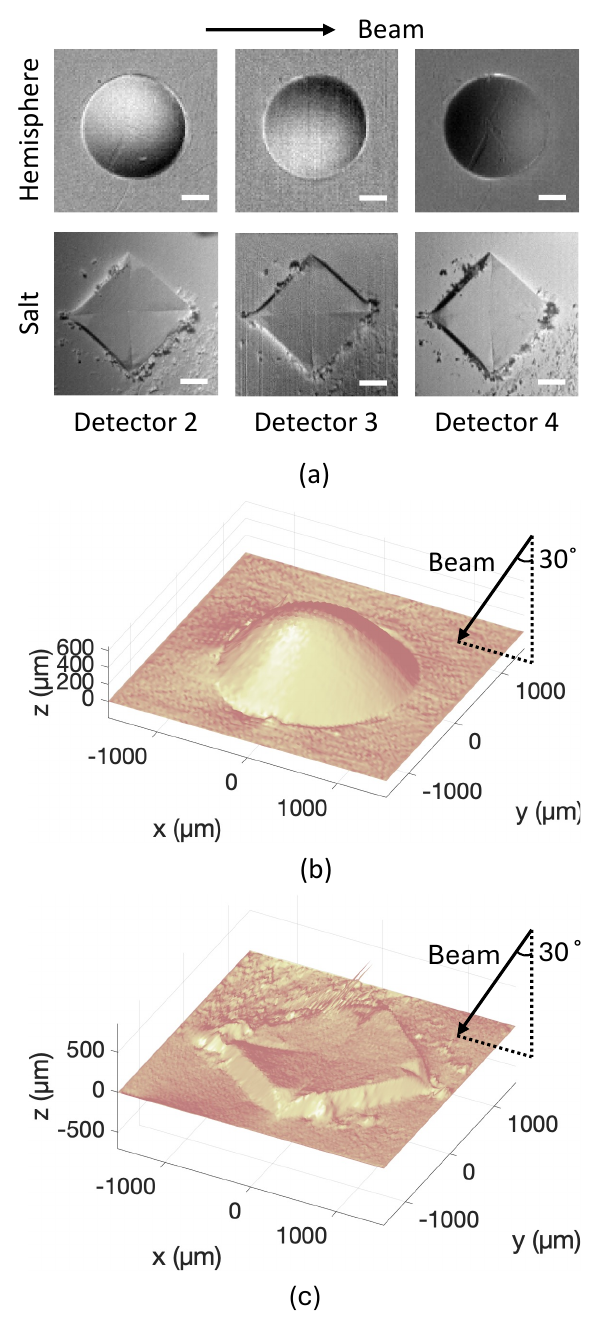}
	\caption{Heliometric stereo reconstructions for an embedded metal sphere and a salt crystal. (a) Micrographs of the hemisphere and the salt crystal captured by three detectors, scale bar $\SI{500}{\micro\metre}$. The helium beam is from the left to the right, as labelled. (b) Reconstructed surfaces of the hemisphere. (c) Reconstruction of the salt crystal. The beam has a 30-degree incidence angle, as indicated in the top right corner of each plot. }
	\label{fig:Reconstructions}
\end{figure}

Two test samples were prepared, first was an embedded metal sphere, which, having a well-defined geometry, allows for a quantification of the errors in the reconstruction; and second a salt crystal. Both samples were imaged in the B-SHeM to obtain three simultaneous images, which were then used to generate a surface profile using the heliometric stereo method. Figure \ref{fig:Reconstructions} shows the acquired SHeM micrographs in panel (a), the reconstruction of the steel embedded sphere in panel (b), and the reconstruction of the salt crystal in panel (c). Qualitatively, both reconstructions work well, recovering the embedded sphere and the detailed, somewhat faceted, surface of the salt crystal. In the case of the salt crystal, a number of quite fine details at the corners of the crustal as well as the four facets of the top face are all recovered, and the reconstruction identifies a rougher region of the substrate (at the top of the reconstruction in figure \ref{fig:Reconstructions}). For the embedded sphere, the overall surface was reconstructed with an 11\% RMS error, with the overall height of the sphere being 94\% of the true height. These errors are slightly larger than those reported in the first implementation of heliometric stereo\cite{radic_3d_2024}, however, in that study 5 micrographs were used to overconstrain the linear system, while here we exactly constrain it, so slightly larger uncertainties are to be expected. The use of multiple detector simultaneously also allows us to significantly reduce the acquisition time compared to approaches where only a single detector is in use. For comparison, the data presented by Radi\'{c} et al.\cite{radic_3d_2024} required a total acquisition time of $\sim15$ hours while the data in the current work, where $\times 2.2$ more pixels were used, the total acquisition time was $\sim 8$ hours. When normalised by the number of pixels and accounting for unused data in the previous work the effective speed-up is $\times4.8$, an impressive increase (the increase is less than would be predicted based on pixel numbers alone as shorter dwell times per pixel were used in the previous work). Like for like comparison cannot be made for the other data presented in the current work, however, it is reasonable to estimate measurement reduction times well in excess of $\times 2$ in sections \ref{sec:multiple_species} and \ref{sec:multiple_modes} once changes of instrument configuration are factor in as well as the simple $\times 2$ rate of data point acquisition. 

\subsection{Multiple species imaging}
\label{sec:multiple_species}

To date, almost all published neutral atom micrographs have used helium as the probe particles, although there is no intrinsic reason preventing the use of other species. Indeed, the two studies that have explored using alternative species, neon/krypton by Witham and Sanchez\cite{witham_exploring_2014} and krypton clusters by Bhardwaj et al.\cite{bhardwaj_contrast_2023} demonstrated not only that imaging with neutral species other than helium is possible, but that there may be alternative contrast mechanisms accessible in that way. The implementation of multiple detectors on the B-SHeM opens up an interesting possibility: simultaneous imaging with two different species using a mixed gas beam. In order to directly compare the micrographs from two species, a pinhole-plate was designed that splits the signal from a single aperture symmetrically into two detectors.

\begin{figure}
	\centering
	\includegraphics[width=\linewidth]{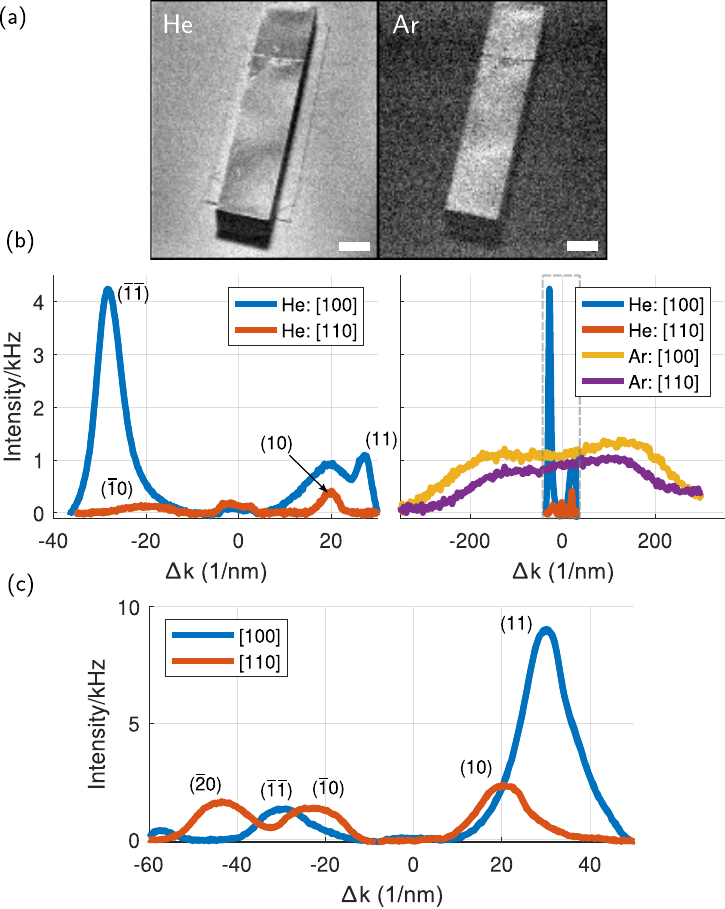}
	\caption{(a) helium and argon images taken of a cleaved flake of LiF, simultaneously acquired with a mixed beam, scale bar $\SI{1}{\milli\metre}$. (b) spot profile diffraction scans taken along (approximately) the two major azimuths with the mixed beam. The left plot is a magnification of the right. (c) a spot profile diffraction scan taken with a pure He beam from the same sample and with the same detection configuration.}
	\label{fig:mixed_gas_diffraction}
\end{figure}

To investigate the potential of mixed gas imaging, a test sample of cleaved LiF was prepared. A mixed beam of 80\% helium and 20\% argon was used to acquire both micrographs and spot profile diffraction scans of the type demonstrated by von Jeinsen et al.\cite{von_jeinsen_2d_2023}. Figure \ref{fig:mixed_gas_diffraction} (a) shows the helium and argon micrographs, where a significant difference in contrast is observed. We can understand the difference in contrast by looking at the spot profile diffraction patterns in (b), where it is clear that with helium we have the momentum resolution to resolve individual diffraction peaks, whereas with argon we do not: thus in the micrograph we are likely off a diffraction peak in helium, rendering the LiF relatively dark, while a significant flux of argon is still scattered, causing the LiF to be rendered relatively light. In terms of measuring diffraction the helium beam clearly has advantages, however, the argon emphasises overall topography more than diffractive features, giving two complementary contrast modes measured simultaneously with the mixed beam. A further advantage of the mixed beam is that it allows modulation of the helium wavelength, in a mixed beam the mixture of helium and argon has a greater effective mass compared to pure helium, leading to a lower terminal velocity and longer wavelength compared to pure helium\cite{scoles_atomic_1988}, the expected velocities and wavelengths of our beam species are given in table \ref{tab:expected parameters}. Thus, using a mixed beam can potentially increase the momentum resolution of helium micro-diffraction measurement, as the same angular range scanned in an experiment corresponds to a smaller momentum range. In our case we, in principle, reduce the instrument broadening of diffraction peaks by 40\% due to our finite angular resolution. Given the angular resolution in micro-diffraction measurements performed in SHeM is generally poorer than more traditional atom scattering measurements\cite{von_jeinsen_2d_2023} wavelength modulation may be crucial in future studies of more complex surface structures.

Figure \ref{fig:mixed_gas_diffraction} plots both the pure (c) and mixed beam (b) helium diffraction measurements. We see the overall range of $\Delta K$ is reduced for the mixed beam due to the increased wavelength, while the angular range of the scans are the same. The FWHM of the $(\bar{1}\bar{1})$ peak for the pure helium was $17.7\pm\SI{0.3}{\nano\metre^{-1}}$ while for the mixed beam it was $9.8\pm\SI{0.2}{\nano\metre^{-1}}$, a reduction of 45\%, similar to the value estimated from the change in helium wavelength (our uncertainties were from the fit and from the uncertainties in the wavelengths). While the increase in wavelength of the helium atoms has yielded us an improved resolution, it has also reduced the range of accessible peaks, the $(\bar{2}0)$ peak for example is accessible with the pure helium but not the mixed, highlighting that the wavelength modulation is a trade-off between resolution and range. In addition, we observe some differences between the relative intensity of the peaks in the two patterns, with the $(\bar{1}\bar{1})$ peak in particular having higher intensity in the mixed gas beam. For some of the other peaks, it is more difficult to draw conclusions, such as the $(11)$ peak which is not fully captured in the mixed beam data case due to the finite range of measurement. We also note the double peak in the mixed gas data around $20-\SI{30}{\nano\metre^{-1}}$, near where the $(11)$ peak would be expected at $\sim\SI{30}{\nano\metre^{-1}}$. While firm explanations of the feature would require further investigations beyond the scope of the current work, we have identified two possibilities. First, is that the sharp minima within the diffraction peak is caused by bound state resonance in the attractive part of the helium-surface potential. Bound state resonances are known to create sharp features within diffraction peaks in helium scattering\cite{riley_refined_2007}, and their appearance will depend on the incident energy of the beam, which would explain why the feature does not appear in the pure helium data. The second, prosaic, explanation, that we cannot exclude, is that there has been a loss of alignment at large positive $\Delta K$, where alignment and momentum resolution are poorest in our experimental setup, and thus diffraction peaks from a different part of hte sample 

\begin{table}[!h]
    \centering
    \begin{tabularx}{\linewidth}{L C C}
        \toprule
        Species       & Velocity/$\SI{}{\metre s^{-1}}$ & Wavelength/$\SI{}{\pico\metre}$\\ \midrule
        Pure He       & $1750\pm20$ & $57.0\pm0.6$ \\
        He in mixture & 1050 & 95 \\
        Ar in mixture & 1050 & 9.5 \\ \bottomrule
    \end{tabularx}
    \caption{The expected velocity and wavelength of the atoms in the pure helium and 80:20 helium:argon mixed beam used to acquire the data in figure \ref{fig:mixed_gas_diffraction}.}
    \label{tab:expected parameters}
\end{table}

\section{Conclusions}

In the present paper, we described our motivation for developing a new, multi-detector helium atom microscope, and our subsequent implementation of the instrument.  We described in detail the guiding principles that we applied, the resulting design, and the steps taken to ensure it can perform optimally under a wide range of operating conditions.  The optimised multi-detector capability opens up a range of new measurements, particularly for 3D surface reconstruction, imaging of transparent or delicate materials, and characterising 2D materials with absolute surface sensitivity.  We presented three case studies illustrating these new capabilities, including simultaneous scattering measurements, topographic reconstruction and mixed gas measurements. For 3D reconstructions, the multi-detector approach provided an effective speed-up in acquisition time of $\times 4.8$ compared to our previous work. While the full benefits of the increased rate of data acquisition have not been explored for other use cases, we expect that the improved throughput of multi-detector SHeMs will significantly increase the scientific productivity of our instrument compared to previous generations. With the mixed gas measurement, we highlighted the possibility of modulating the helium wavelength to change the momentum resolution and range for micro-diffraction measurement. We hope these developments will make SHeM measurements easier and will stimulate new and exciting applications. Our new instrument will be available for access through \textsc{corde} (\url{https://corde.phy.cam.ac.uk}), so we look forward to wider community engagement in order to address a wide range of scientific challenges.

\section*{Supporting data}

A data-pack supporting this publication can be found at \url{doi.org/placeholder}. \emph{On final submission a link to the supporting data will be inserted.}

\section*{Acknowledgments}
	
The work was supported by \textsc{epsrc} grant EP/R008272/1. The work was performed in part at \textsc{corde}, the Collaborative R\&D Environment established to provide access to physics related facilities at the Cavendish Laboratory, University of Cambridge and \textsc{epsrc} award EP/T00634X/1. SML acknowledges support from IAA award EP/X525686/1. The authors acknowledge support from Ionoptika Ltd.

\bibliographystyle{unsrtnat}
\bibliography{multi_detector.bib}

\end{document}